\documentclass[a4paper,11pt]{article}
\usepackage{amsthm,graphicx,latexsym,amsmath,amssymb} 
\usepackage[left=2cm,right=2cm,top=2cm,bottom=2cm]{geometry}

\graphicspath{{eps/}}

\newcommand{\nn}{\nonumber}

\newcommand{\myfig}[2]{
\begin{figure}
  \begin{center}
    \includegraphics{#1.eps}
    \caption{#2}
    \label{fig:#1}
  \end{center}
\end{figure}
}
\newcommand{\figref}[1]{Fig.~\ref{fig:#1}}

\newcommand{\On}{{\rm O}(n)}
\newcommand{\half}{\frac{1}{2}}
\newcommand{\wh}{\widehat}
\newcommand{\ZN}{\mathbb{Z}_N}

\begin{document}

\title{Discretely holomorphic parafermions and integrable boundary conditions}

\author{Yacine Ikhlef}

\maketitle

\abstract{
  

  In two-dimensional statistical models possessing a discretely holomorphic
  parafermion, we introduce a modified discrete Cauchy-Riemann equation
  on the boundary of the domain, and we show that the solution of this
  equation yields integrable boundary Boltzmann weights.
  This approach is applied
  to (i) the square-lattice $\On$ loop model, where the exact locations
  of the special and ordinary transitions are recovered, and (ii) the 
  Fateev-Zamolodchikov $\ZN$ spin model, where a new
  rotation-invariant, integrable boundary condition is discovered for
  generic $N$.
}

\section{Introduction}

Discretely holomorphic parafermions (also called lattice parafermions) 
are lattice analogs for
the holomorphic fields of Conformal Field Theory (CFT).
They are lattice observables which satisfy a discrete version
of the Cauchy-Riemann (CR) equation, and
were recently identified in various
two-dimensional critical lattice models, which are either spin
models with a local interaction or loop models describing
self-avoiding walks (SAWs) or extended interfaces.
Generally, in spin models with Kramers-Wannier duality (see~\cite{Dubedat11}
for a recent review),
a lattice parafermion is constructed~\cite{RivaC06,RajC07} by taking the
product of lattice spin and disorder operators, providing an exact lattice
analog for the operator product expansion (OPE) of the corresponding 
CFT~\cite{ZamoloF85}.
In loop models, the lattice parafermion is expressed~\cite{RivaC06,IkhlefC09} in terms
of a single open path attached to the boundary of the system.

Lattice parafermions have found many applications to 
the rigorous study of lattice models (especially the Ising and SAW models)
in the scaling limit, providing
mathematical proofs for the exact results obtained by the Coulomb gas
approach~\cite{Nienhuis84,DFSZ87} thirty years ago.
In particular, the lattice parafermion is a
crucial ingredient in the proofs~\cite{ChelkakS09,HonglerK11} of convergence of
Ising domain walls to the Schramm-Loewner Evolution (SLE), and in the
calculation of Ising correlation functions~\cite{ChelkakHI12}.
Also, for the SAW model on the honeycomb lattice,
the lattice parafermion is essential to determine rigorously the location
of the critical weights, both in the bulk~\cite{DuminilS10} and
on the boundary~\cite{BeatondGG11,BeatonGJ12}.

This paper is concerned with lattice models at finite size,
and addresses the relation between lattice parafermions
and quantum integrability. Previous studies~\cite{RivaC06,RajC07,IkhlefC09}
have shown that many (spin or loop)
statistical models admit a lattice parafermion exactly at their integrable
point. In other words, the Boltzmann weights $\{W(\alpha)\}$ which satisfy
the discrete CR equations on the rhombic lattice of angle $\alpha$
are also a solution of the Yang-Baxter (YB) equations, where
$\alpha$ plays the role of the spectral parameter.
In the present work, it is shown that a similar relation
holds between the solutions of CR equations {\it on the boundary}
and of Sklyanin's reflection equation~\cite{Skl88} (or boundary YB equation).

Two distinct models are considered in this paper. First, for the
$\On$ loop model on the square lattice~\cite{Nien90}, following
some ideas used in~\cite{BeatondGG11},
local CR equations around boundary faces are introduced, and their solution
is related to the boundary integrable weights found in~\cite{YungB95}. When the
rhombus angle is set to $\alpha=\pi/3$, one recovers the result
of~\cite{BeatondGG11} for the honeycomb lattice. Then, we study the 
Fateev-Zamolodchikov $\ZN$ spin model~\cite{FateevZ82}.
We introduce a similar 
boundary CR equation, and show that its solution yields new non-trivial
integrable weights, associated to {free} BCs.
In both cases, we use the lattice parafermions constructed
in~\cite{RajC07,IkhlefC09} which satisfy bulk CR equations,
and we define a new type of lattice
domain $\Omega$ where some of the boundary faces have a differnet shape from
the bulk ones.

\section{The $\On$ model on the square lattice}

\subsection{Integrable weights and Cauchy-Riemann equation in the bulk}

\myfig{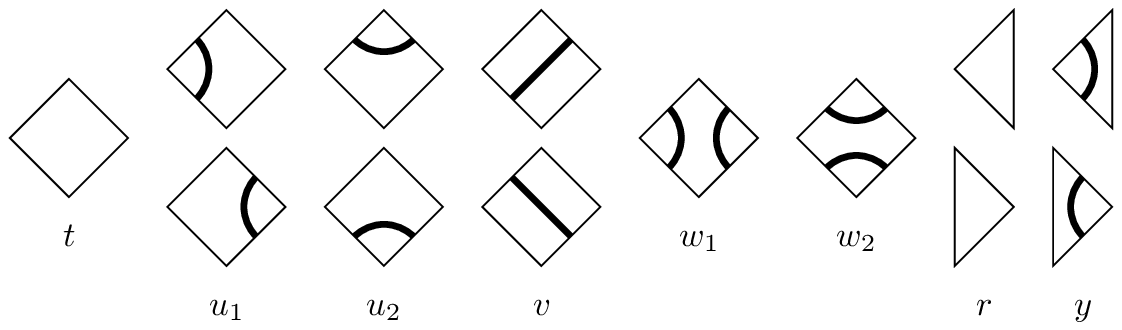}{The local configurations of the square-lattice $\On$ model and the
  associated Boltzmann weights.}

\myfig{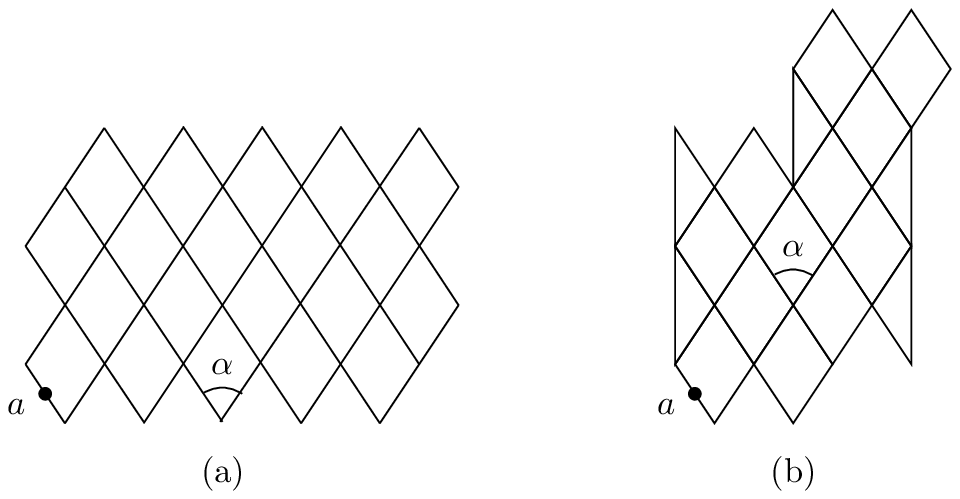}{(a) A domain $\Omega_0$ of the rhombic lattice with angle
  $\alpha$. (b) A domain $\Omega$ of the rhombic lattice, where some boundary
  faces are replaced by triangles. In both cases, $a$ is an arbitrary midedge
  of a boundary rhombus.}

The $\On$ model on the square lattice was introduced in~\cite{Nien90}.
It is a loop model, defined by the local configurations shown in \figref{On},
and the Boltzmann weight of a loop configuration $G$:
$$
\Pi[G] = n^{N_\ell(G)} \times t^{N_t(G)} u_1^{N_{u_1}(G)} \dots w_2^{N_{w_2}(G)} \,,
$$
where $N_\ell(G)$ is the number of closed loops in $G$, and
$N_x$ is the number of faces with configuration $x$. The partition function
is denoted by $Z$. If the loop fugacity is set to
$$
n = -2 \cos 4\lambda \,,
$$
the integrable weights are given by
\begin{equation} \label{eq:weights-On}
  \begin{array}{rcl}
    t &=& \sin(3\lambda-u)\sin u + \sin 2\lambda \sin 3\lambda \\
    u_1 &=& -\sin(3\lambda-u) \sin 2\lambda \\
    u_2 &=& -\sin u \sin 2\lambda \\
    v &=& \sin(3\lambda-u) \sin u \\
    w_1 &=& \sin(3\lambda-u) \sin(2\lambda-u) \\
    w_2 &=& -\sin(\lambda-u) \sin u \,,
  \end{array}
\end{equation}
where $u$ is the spectral parameter. We first consider
a domain $\Omega_0$ of the rhombic lattice ${\cal R}_\alpha$ (see \figref{domain}a),
and recall the results on the lattice parafermion for the $\On$ model.
For a fixed edge $a$ on the boundary, the parafermionic observable $F_s$
is defined on the midedges $\{z\}$ of the rhombi as~\cite{IkhlefC09}
\begin{equation} \label{eq:Fs-On}
  F_s(z) := \frac{1}{Z} \sum_{G \in \Gamma(a\to z)} \Pi[G] \ e^{-is \theta(G)} \,,
\end{equation}
where $\Gamma(a\to z)$ is the set of loop configurations comprising an open
path between $a$ and $z$, $\theta(G)$ is the winding angle of the
path included in $G$, and $s$ is called the conformal spin.
For a polygon $P$ with vertices given by
the complex numbers $(p_1, \dots,p_m)$ ordered counter-clockwise, we define
the discrete contour integral of a function $f$ around $P$ as
\begin{equation} \label{eq:latt-integral}
  \sum_P f(z) \delta z :=
  \sum_{j=1}^m (p_{j+1}-p_j) f \left(\frac{p_{j+1}+p_j}{2}\right) \,,
\end{equation}
where $p_{m+1}:=p_1$. We define the discrete CR equations as
\begin{equation}
  \sum_{\displaystyle \diamond} F_s(z) \delta z = 0 \,,
\end{equation}
where the left-hand side stands for the discrete integral around any given
rhombus of $\Omega_0$.
Fixing the outer loop configuration, this gives the linear system:
\begin{equation} \label{eq:CR-On}
  \begin{array}{rcl}
    t + \mu u_1 - \mu\tau^{-1} u_2 - v &=& 0 \\
    -\tau^{-1} u_1 + nu_2 + \tau\mu v -\mu\tau^{-1}(w_1+nw_2) &=& 0 \\
    nu_1- \tau u_2 - \mu\tau^{-2} v + \mu(nw_1 + w_2) &=& 0 \\
    -\mu\tau^{-2} u_1 + \mu\tau u_2 +nv-\tau^{-2} w_1-\tau^2 w_2 &=& 0 \,,
  \end{array}
\end{equation}
where we have set
$$
  \tau := e^{i\pi s} \,,
  \qquad
  \mu := e^{i(s+1) \alpha} \,.
$$
In~\cite{IkhlefC09} it was shown that the system~\eqref{eq:CR-On} has a nonzero solution iff
$\cos 4\pi s = \cos 12\lambda$, and one can choose the value
\begin{equation} \label{eq:s-On}
  s = \frac{3\lambda}{\pi}-1 \,.
\end{equation}
Moreover, the solution of~\eqref{eq:CR-On} for $v\neq 0$
is given by the integrable weights~\eqref{eq:weights-On}, if one sets the
spectral parameter to
\begin{equation} \label{eq:u-On}
  u = (s+1)\alpha = \frac{3\lambda}{\pi} \ \alpha \,.
\end{equation}

\subsection{Boundary Cauchy-Riemann equation}

We now consider a domain $\Omega$ where the interior faces are rhombi
with a lower angle~$\alpha$, and the boundary faces are either rhombi
or triangles (see~\figref{domain}b). Moreover, we assume that the initial
orientation of the open path at the boundary point
$a$ is horizontal. The Boltzmann weights on a triangular face are $y,r$
(see~\figref{On}).

\myfig{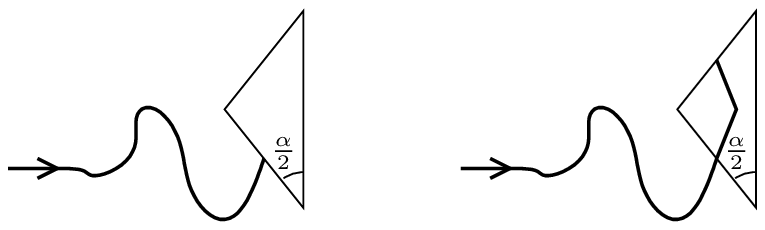}{The two possible configurations of a triangular face contributing
  to~\eqref{eq:bCR-On} for a fixed outer configuration.}

Let us introduce the boundary CR equation:
\begin{equation} \label{eq:bCR-On}
  {\rm Re} \left[
    \sum_{\displaystyle \triangleleft} F_s(z) \delta z
    \right] = 0 \,,
\end{equation}
where the discrete contour integral is around a triangular boundary face $T$.
The configurations $G$ in~\eqref{eq:Fs-On} which contribute
to~\eqref{eq:bCR-On} are those where the open path ends on an edge
of the boundary face $T$. Like in the bulk, one can fix the loop
configuration outside $T$ and sum over the internal configurations
(see~\figref{holo}).
This yields
\begin{equation} \label{eq:holo2}
  {\rm Re} \left[
    -e^{\frac{i}{2}[(s+1)\alpha+\pi(1-s)]} r
    -e^{\frac{i}{2}[-(s+1)\alpha+\pi(1-s)]} y
    \right] = 0 \,.
\end{equation}

We now have to choose an appropriate parameterisation for the spectral
parameter. For this, we consider a particular instance of a domain $\Omega$:
the infinite strip in a diagonal direction of ${\cal
  R}_\alpha$. In~\cite{YungB95}, it is shown how to construct an integrable
model on a strip with arbitrary, homogeneous spectral parameter in the bulk. 
If we denote $\check{R}$ and $K$ the solutions to the bulk and boundary YB equations,
the resulting model has bulk weights given by $\check{R}(u)$, and boundary weights
given by $K(u/2)$. This suggests that in~\eqref{eq:holo2}, one should set
$$
u = \half (s+1)\alpha
$$
instead of~\eqref{eq:u-On}. In contrast, the conformal spin belongs to the
definition of the observable $F_s$, and we keep its value~\eqref{eq:s-On}.
The solution of~\eqref{eq:holo2} then reads
\begin{equation} \label{eq:K-On}
  \begin{array}{rcl}
    r(u) &=& -\cos \half(3\lambda+2u) \\
    \\
    y(u) &=& \phantom{-} \cos \half(3\lambda-2u) \,.
  \end{array}
\end{equation}
These are exactly the boundary weights identifed in~\cite{YungB95},
which satisfy the boundary YB equation
$$
\check{R}_{12}(u-v) K_2(u) \check{R}_{12}(u+v) K_2(v)
= K_2(v) \check{R}_{12}(u+v) K_2(u) \check{R}_{12}(u-v) \,.
$$
In~\cite{YungB95}, it is argued that~\eqref{eq:K-On}
corresponds to the special surface transition in a certain regime of $\lambda$.
Under the change $\lambda \to \lambda+\pi$, the bulk weights are invariant
(up to irrelevant sign factors), whereas the boundary weights correspond to
the ordinary surface transition.

Note that, if we specialise to $u=\lambda$ ($\alpha=\frac{\pi}{3}$),
we recover the critical bulk and boundary weights of the $\On$ model
on the honeycomb lattice, which were found through a type of boundary CR
equation in \cite{BeatondGG11} for a specific domain $\Omega$.

\section{The Fateev-Zamolodchikov $\ZN$ model}

\subsection{Definitions}

\myfig{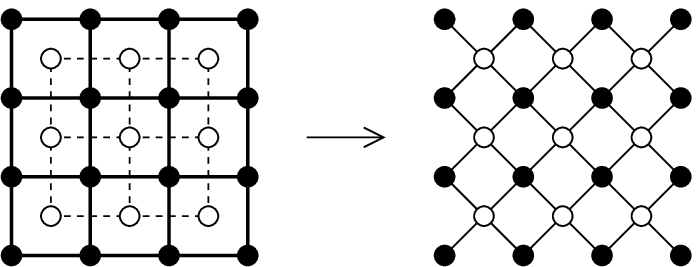}{Left: the original square lattice $\cal L$ for the $\ZN$
  spins $\sigma_j$ (full dots, full lines)
  and its dual lattice (empty dots, dotted lines). Right: the covering
  lattice ${\cal R}$.}

The $\ZN$ model consists of spins $\sigma_j$ living on the sites of
the square lattice $\cal L$ (full dots in \figref{covering}), and taking the values:
\begin{equation}
  \sigma_j \in \{ 1, \omega, \dots, \omega^{N-1}\} \,,
  \qquad \text{where} \qquad
  \omega=\exp\left(\frac{2i\pi}{N}\right) \,.
\end{equation}
We shall then write $\sigma_j=\omega^{r_j}$, with $r_j \in \mathbb{Z}$.
We also introduce the crossing parameter
$$
\lambda := \frac{\pi}{2N} \,.
$$
The Boltzmann weight of a spin configuration is:
\begin{equation}
  \Pi[\sigma] = \prod_{\langle ij \rangle} W(\sigma_i,\sigma_j) \,,
  \qquad \text{where} \qquad
  W(\sigma_i,\sigma_j) = \begin{cases}
    W(u|r_i-r_j) & \text{if $\langle ij \rangle$ is horizontal,} \\
    W(\lambda-u|r_i-r_j) & \text{if $\langle ij \rangle$ is vertical,}
  \end{cases}
\end{equation}
where $u$ is the spectral parameter. Moreover, we restrict to a
periodic and symmetric function $W(u|r)$:
\begin{equation} \label{eq:sym-W}
  W(u|r+N) = W(u|-r) = W(u|r) \,.
\end{equation}
The partition function is then defined as
\begin{equation}
  Z = \sum_{\{\sigma\}} \Pi[\sigma] \,,
\end{equation}
and a typical correlation function of $n$ spin operators is of the form
$$
\langle \sigma_{j_1}^{k_1} \sigma_{j_2}^{k_2} \dots \sigma_{j_n}^{k_n} \rangle
:= \frac{1}{Z} \sum_{\{\sigma\}} \Pi[\sigma] 
\sigma_{j_1}^{m_1} \sigma_{j_2}^{m_2} \dots \sigma_{j_n}^{m_n} \,,
$$
where $\{j_1, \dots, j_n\}$ are $n$ sites of the lattice $\cal L$, and 
$\{m_1, \dots, m_n\}$ are $n$ arbitrary integers. As a consequence of the
symmetries~\eqref{eq:sym-W} of the Boltzmann weights, the model enjoys
Kramers-Wannier duality, and in particular one may define dual spin
operators, also called disorder operators $\mu^{(m)}$. These
live on the dual lattice (empty dots in
\figref{covering}), and are defined through their correlation functions.
For the purposes of the present study, it is sufficient to define
correlation functions involving two $\mu$ operators.
Let $k$ and $\ell$ be sites of the dual lattice ${\cal L}'$,
and $\gamma_{k\ell}$ an oriented path
from $k$ to $\ell$ on ${\cal L}'$.
One then defines:
\begin{equation}
  \langle \mu_k^{(-m)} \mu_{\ell}^{(m)} (\dots) \rangle :=
  \frac{1}{Z} \sum_{\{\sigma\}}
  \prod_{\langle ij \rangle \notin \gamma_{k\ell}^\perp} W(\sigma_i,\sigma_j)
  \prod_{\langle ij \rangle \in \gamma_{k\ell}^\perp}
  W(\sigma_i,\omega^m\sigma_j) \ (\dots)\,,
\end{equation}
where the dots denote any function of the spins,
$\gamma_{k\ell}^\perp$ is the set of edges of $\cal L$ which cross $\gamma_{k\ell}$,
and we have taken the convention that $\sigma_i$ ($\sigma_j$) is on the left (right) of
$\gamma_{k\ell}$. The symmetries~\eqref{eq:sym-W} imply that
$\langle \mu_k^{(-m)} \mu_\ell^{(m)} \rangle$ is independent of the
choice of the path $\gamma_{k\ell}$.

In the following, we shall consider a deformation of the square lattice $\cal
L$ into a rectangular lattice of aspect ratio $\tan \frac{\alpha}{2}$. The
union of $\cal L$ and its dual is then a rhombic lattice of angle $\alpha$, 
called the covering lattice and denoted ${\cal R}_\alpha$.

Later in this Section, we will use the discrete Fourier transform
\begin{equation}
  \widehat{f}(k) := \sum_{r=0}^{N-1} \omega^{kr} f(r) \,,
  \qquad
  f(r) := \frac{1}{N} \sum_{r=0}^{N-1} \omega^{-kr} \widehat{f}(k) \,.
\end{equation}

\subsection{Integrable weights and Cauchy-Riemann equation in the bulk}

\myfig{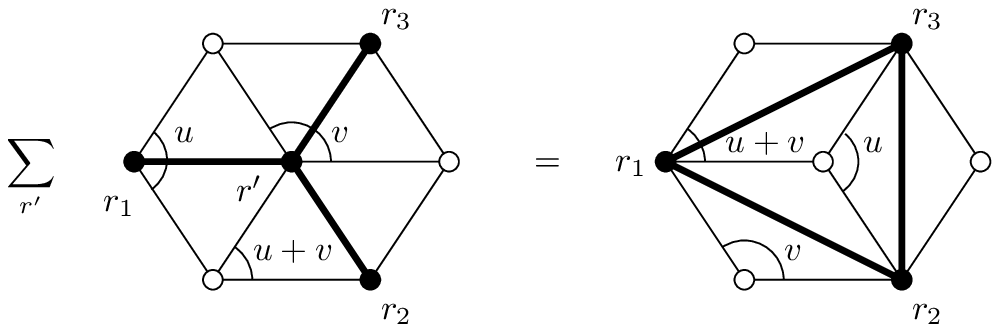}{The star-triangle relation as a special case of the
  Yang-Baxter equation.}

The integrability condition for a periodic system ({\it i.e.} the commutation
of transfer matrices on the cylinder) has the form of a star-triangle
equation
\begin{eqnarray}
  \sum_{r'=0}^{N-1} W(\lambda-u|r_1-r') W(u+v|r_2-r') W(\lambda-v|r_3-r') \nn \\
  = W(u|r_2-r_3) W(\lambda-u-v|r_1-r_3) W(v|r_1-r_2) \,,
\end{eqnarray}
which can be formulated as the Yang-Baxter equation on the covering lattice
(see \figref{star-triangle}), and a solution to this equation was given
in~\cite{FateevZ82}:
\begin{equation} \label{eq:FZ}
  W(u|0) = 1 \,,
  \qquad
  W(u|r) = \prod_{p=0}^{r-1}
  \frac{\sin[(2p+1)\lambda-u]}{\sin[(2p+1)\lambda+u]}
  \quad \text{for $r \in \{1, \dots, N-1 \}$.}
\end{equation}
This solution is self-dual, in the sense that
\begin{equation} \label{eq:W-sd}
  \wh{W}(u|k) = W(\lambda-u|k) \,.
\end{equation}

\myfig{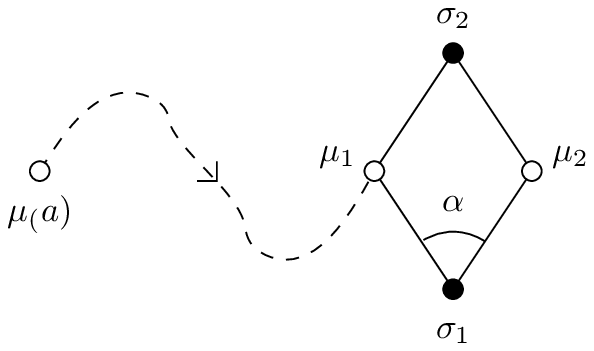}{A face of the rhombic covering lattice ${\cal R}_\alpha$
  in the $\ZN$ model. The dotted line represents an arbitray path
  $\gamma$ on ${\cal L}'$ from $\mu(a)$ to $\mu_1$.}

The construction~\cite{RajC07} of the discrete parafermion in the $\ZN$ model
is the lattice version of the parafermionic current in the
$\ZN$ CFT~\cite{ZamoloF85}. In the lattice model, as well as in the
CFT, the parafermion $\psi$ is obtained as the product of a spin
operator~$\sigma$ and a disorder operator~$\mu$ on adjacent sites.
Consider a point $z$ on the middle of an edge of ${\cal R}_\alpha$,
and denote $j(z),k(z)$ the sites of ${\cal L},{\cal L}'$ adjacent to
$z$. We shall use the short-hand notations
$$
\sigma(z) := \sigma_{j(z)} \,,
\qquad
\mu(z) := \mu^{(1)}_{k(z)} \,,
\qquad
\mu^*(z) := \mu^{(-1)}_{k(z)} \,.
$$
The lattice parafermion $\psi(z)$ is defined as~\cite{RajC07}
\begin{equation} \label{eq:psi}
  \psi(z) := \sigma(z) \ \mu(z) \,.
\end{equation}
The observable $F_s(z)$ is then defined by fixing a point $a$ on the
boundary, and setting
\begin{eqnarray}
  F_s(z) &:=& \langle e^{-is\theta(a,z)} \psi^*(a) \psi(z) \rangle \nn \\
  &=& \langle e^{-is\theta(a,z)} \sigma^*(a) \sigma(z)
  \mu^*(a) \mu(z) \rangle \,,
\end{eqnarray}
where $\theta(a,z)$ is the angle between the vectors $\overrightarrow{k(a)j(a)}$
and $\overrightarrow{k(z)j(z)}$, and $s$ is the conformal spin. 
As in the $\On$ model, one defines the discrete CR equation on ${\cal R}_\alpha$
as
\begin{equation} \label{eq:CR-ZN1}
  \sum_{\displaystyle \diamond} F_s(z) \delta z = 0 \,,
\end{equation}
which can be written (see \figref{plaqZN})
\begin{equation} \label{eq:CR-ZN2}
  e^{-s\theta(a)} \ 
  \langle \sigma^*(a) \sigma_1 \mu^*(a) \mu_1 \left[
    (e^{-iu} 
    - e^{iu} \sigma_1^*\sigma_2)
    + (e^{iu + i\pi s}
    - e^{-iu - i\pi s} \sigma_1^* \sigma_2) \mu_1^* \mu_2
    \right] \rangle = 0 \,,
\end{equation}
where $\theta(a)$ is the angle between $\overrightarrow{k(a)j(a)}$ and the
horizontal, and we have set
\begin{equation} \label{eq:u-ZN}
  u = \half (1-s)(\pi-\alpha) \,.
\end{equation}
From the definition of $\mu$, \eqref{eq:CR-ZN2} reads
\begin{eqnarray}
  &&\frac{1}{Z} \sum_{\{\sigma\}} \sigma^*(a)\sigma_1
  \prod_{\langle ij \rangle \notin \gamma^\perp \cup \{\langle 12 \rangle\}}
  W(\sigma_i,\sigma_j)
  \prod_{\langle ij \rangle \in \gamma^\perp}
  W(\sigma_i,\omega\sigma_j)  \nn \\
  && \qquad \times \left[
    (e^{-iu} - e^{iu} \sigma_1^*\sigma_2) W(\sigma_1,\sigma_2)
    + (e^{iu + i\pi s} - e^{-iu - i\pi s}\sigma_1^*\sigma_2) W(\sigma_1,\omega\sigma_2)
    \right] = 0 \,,
  \label{eq:CR-ZN3}
\end{eqnarray}
where $\gamma$ is an arbitrary path on ${\cal L}'$ going from $\mu(a)$ to
$\mu_1$ (see~\figref{plaqZN}).
The bracket in~\eqref{eq:CR-ZN3} may be written as
\begin{equation}
  I(r) := (e^{-iu} - e^{iu} \omega^r) W(u|r)
    + (e^{iu + i\pi s} - e^{-iu - i\pi s} \omega^r) W(u|r+1) \,,
\end{equation}
where
$$
\sigma_1^* \sigma_2 = \omega^r \,.
$$
Therefore, a sufficient condition for~\eqref{eq:CR-ZN2} to hold is that
$I(r)$ vanishes for all $r$, which yields the recursion relation
\begin{equation} \label{eq:rec-W}
  W(u|r+1) = W(u|r) \times (-e^{i\pi s+\frac{i\pi}{N}})
  \ \frac{\sin \left[\frac{(2r+1)\pi}{2N}-u \right]}
       {\sin \left[\frac{(2r+1)\pi}{2N}+u \right]} \,.
\end{equation}
Compatibility of~\eqref{eq:rec-W} with the symmetries~\eqref{eq:sym-W} fixes
the value of the spin:
\begin{equation}
  s = 1- \frac{1}{N} \,,
\end{equation}
and for this value, the solution of \eqref{eq:rec-W} is given by
the integrable weights~\eqref{eq:FZ}.

As explained in~\cite{RajC07}, each of the $(N-1)$ discrete parafermions
$\psi^{(m)}(z) = \sigma^m(z) \mu^{(m)}(z)$ with charge
$m \in \{1,\dots, N-1\}$ satisfies the CR equations, but
for a different function $W^{(m)}$, and a conformal spin $s_m=m(N-m)/N$.
In fact, $W^{(m)}$ is simply related
to $W$~\eqref{eq:FZ} by the global transformation $\sigma_j \to \sigma_j^m$.
Hence, one can restrict to the case
$m=1$ exposed above, without loss of generality.

\subsection{Boundary Yang-Baxter equation}

\myfig{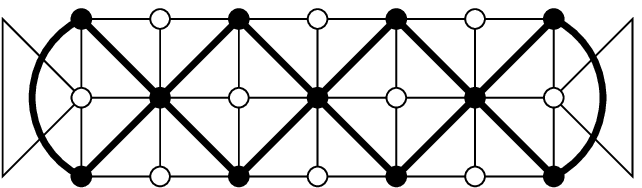}{The two-row transfer matrix along one direction of
  the covering lattice, with open boundary conditions.
  Thick lines denote interaction terms between the spins.}

\myfig{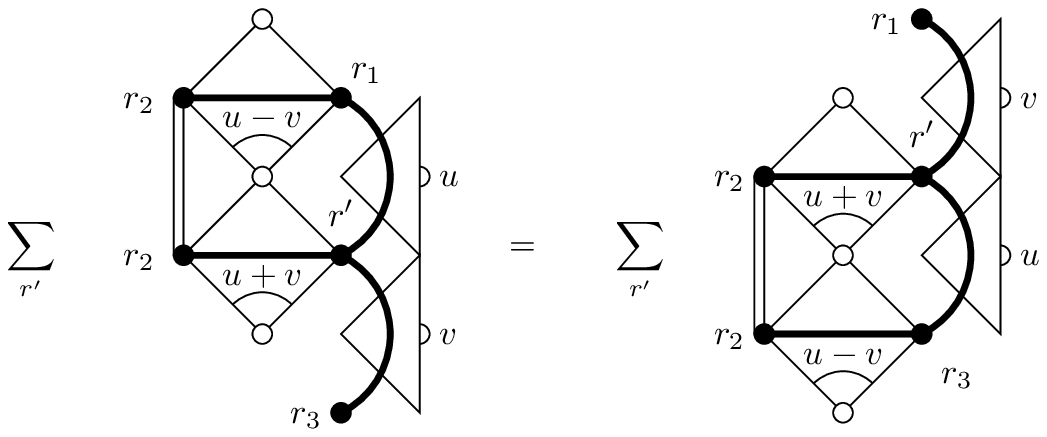}{The boundary Yang-Baxter equation for the $\ZN$ model. Thick lines
  denote interaction between the spins, and spins joined by a double line
  must be equal.}

Following~\cite{Zhou97}, we consider the boundary conditions which
ensure the commutation of two-row transfer matrices
(see \figref{tm-ZN}).
We denote $Y_{R,L}(\sigma_i,\sigma_j)$ the Boltzmann weight
on the right and left boundaries, and the partition function on the strip reads
\begin{equation}
  Z_{\rm strip} = \sum_{\{\sigma\}}
  \prod_{\langle ij \rangle \in {\rm bulk}} W(\sigma_i,\sigma_j)
  \prod_{\langle ij \rangle \in {\rm right \ bound.}} Y_R(\sigma_i,\sigma_j)
  \prod_{\langle ij \rangle \in {\rm left \ bound.}} Y_L(\sigma_i,\sigma_j) \,.
\end{equation}
In~\cite{Zhou97}, fixed-spin integrable boundary conditions ({\it i.e.}
$Y(\sigma_i,\sigma_j)=\delta_{r_i,\alpha} \delta_{r_j,\beta}$ with $\alpha,\beta$ fixed)
were considered, and the associated surface free energies were obtained.
In the present work, we address the boundary YB equation for
rotation-invariant boundary conditions, {\it i.e.} an interaction
of the form
\begin{equation}
  Y(\sigma_i,\sigma_j) = Y(u,\xi|r_i-r_j) \,,
\end{equation}
where $\xi$ is a complex parameter associated to each boundary ({\it boundary field}),
and $Y$ satisfies the symmetries~\eqref{eq:sym-W}.
The boundary YB equation then reads (see \figref{refl-ZN})
\begin{eqnarray}
  W(u-v|r_1-r_2) \sum_{r'=0}^{N-1} Y(u,\xi|r_1-r') W(u+v|r_2-r') Y(v,\xi|r_3-r')
  \nn \\
  =  W(u-v|r_2-r_3) \sum_{r'=0}^{N-1} Y(v|r_1-r') W(u+v|r_2-r')
  Y(u,\xi|r_3-r') \,,
\end{eqnarray}
or, in Fourier space:
\begin{eqnarray}
  \wh{Y}(v,\xi|k) \ \sum_{m=0}^{N-1} \wh{W}(u-v|m) \wh{Y}(u,\xi|\ell-m) \wh{W}(u+v|k+\ell-m) 
  \nn \\
  =  \wh{Y}(v,\xi|\ell) \ \sum_{m=0}^{N-1} \wh{W}(u-v|m) \wh{W}(u+v|k+\ell-m)
  \wh{Y}(u,\xi|k-m) \,.
  \label{eq:refl-ZN}
\end{eqnarray}
These functional equations for $Y$ are difficult to solve for general $N$,
but we shall now show that a nontrivial solution can be found by use
of the boundary CR equation.

\subsection{Boundary Cauchy-Riemann equation}

\myfig{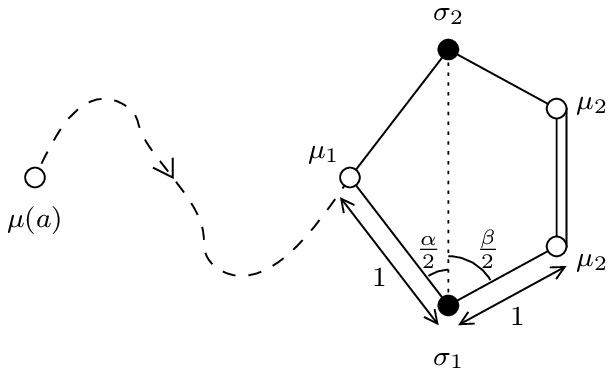}{A boundary face in the $\ZN$ model. Like in~\figref{plaqZN},
  the dotted line represents an arbitray path
  $\gamma$ on ${\cal L}'$ from $\mu(a)$ to $\mu_1$.
  In the discrete contour
  integral~\eqref{eq:int-F}, only the edges represented by single lines
  contribute.}

We consider boundary faces in the shape of a pentagon, determined by the
angles $\alpha$ and $\beta$ (see \figref{kite}).
The contour integral of $F_s$ around such a face $P$ reads:
\begin{equation} \label{eq:int-F}
  \sum_P F_s(z) \delta z = e^{-s\theta(a)} \ 
  \langle \psi^*(a) \sigma_1 \mu_1
  \left[
    (e^{-2iu} 
    - e^{2iu} \sigma_1^*\sigma_2)
    + (e^{i\xi + i\pi s} 
    - e^{-i\xi - i\pi s} \sigma_1^*\sigma_2) \mu_1^* \mu_2
    \right] \rangle\,,
\end{equation}
where we have set\footnote{Like for the $\On$ model, for a given angle
  $\alpha$, the spectral parameter for boundary weights is half the bulk
  value~\eqref{eq:u-ZN}.}
\begin{equation}
  u = \frac{1}{4}(1-s)(\pi-\alpha) \,,
  \qquad \text{and} \qquad
  \xi =  \half(1-s)(\pi-\beta) \,.
\end{equation}
Like for the bulk CR equation, using the notation $\sigma_1^*
\sigma_2^{\phantom{*}} := \omega^r$, \eqref{eq:int-F} is rewritten as
$$
   \sum_P F_s(z) \delta z = \frac{e^{-s\theta(a)}}{Z} \sum_{\{\sigma\}}
   \sigma^*(a)\sigma_1
  \prod_{\langle ij \rangle \notin \gamma^\perp}
  W(\sigma_i,\sigma_j)
  \prod_{\langle ij \rangle \in \gamma^\perp}
  W(\sigma_i,\omega\sigma_j)
  \times J(r) \,,
$$
where $J(r)$ is defined as
\begin{equation}
  J(r):=(e^{-2iu} 
  - e^{2iu} \omega^r) Y(u,\xi|r)
  + (e^{i\xi + i\pi s}
  - e^{-i\xi - i\pi s} \omega^r) Y(u,\xi|r+1) \,,
\end{equation}
and its Fourier transform is
\begin{equation}
  \wh{J}(k):=(e^{-2iu} +e^{i\xi + i\pi s} \omega^{-k}) \wh{Y}(u,\xi|k)
  - (e^{2iu} + e^{-i\xi - i\pi s} \omega^{-k-1}) \wh{Y}(u,\xi|k+1) \,.
\end{equation}
In analogy with the $\On$ model, we introduce a boundary CR equation
of the form
\begin{equation} \label{eq:bCR-ZN}
  {\rm Re} \left[ e^{i\varphi(k)} \wh{J}(k) \right] = 0 \,,
\end{equation}
where $\varphi(k)$ is real.
It turns out that $Y$ only satisfies the symmetry relations~\eqref{eq:sym-W}
for the choice $\varphi(k)=(4k+2)\lambda$, and the solution
of~\eqref{eq:bCR-ZN} then reads
\begin{equation} \label{eq:Y}
  \wh{Y}(u,\xi|k) = \begin{cases}
    1 & \text{for $k=0$.} \\
    {\displaystyle \prod_{\ell=0}^{k-1}
    \frac{\sin\left[ (2\ell+1)\lambda-u+\xi \right]
      \sin\left[ (2\ell+1)\lambda-u-\xi \right]}
         {\sin\left[ (2\ell+1)\lambda+u+\xi \right]
           \sin\left[ (2\ell+1)\lambda+u-\xi \right]}}
         & \text{for $k \in \{1, \dots, N-1\}$.}
  \end{cases}
\end{equation}
A remarkable fact is that {\it the weights~\eqref{eq:Y} are also a
solution of the boundary YB equation~\eqref{eq:refl-ZN}}.
The solutions $W^{(m)},Y^{(m)}$ associated to the other
parafermions $\psi^{(m)}$ also satisfy the boundary YB equation,
and they are related to \eqref{eq:FZ}--\eqref{eq:Y} by a simple transformation
$\sigma_j \to \sigma_j^m$.

\subsection{Physical interpretation of the integrable boundary condition}

To indentify the BC corresponding to~\eqref{eq:Y}, we consider the
{\it diagonal-to-diagonal} transfer matrix, which is constructed from the
two-row transfer matrix by the procedure described in~\cite{YungB95}.
If the two-row transfer matrix with horizontal spectral parameter $u$,
vertical spectral parameters $(v_1, \dots, v_N)$ and boundary parameters
$\xi_{L,R}$ is denoted
$$
t(u,\xi_L,\xi_R|v_1, \dots, v_N) \,,
$$
then the diagonal-to-diagonal transfer matrix is given by
$$
t_d (u,\xi_L,\xi_R) := t\left(
\frac{u}{2},\xi_L,\xi_R \right|\left. \frac{u}{2},-\frac{u}{2},
\dots, \frac{u}{2},-\frac{u}{2}
\right) \,.
$$
In this setting, the bulk weights are given by $W(u|\cdot)$ for horizontal
edges and $W(\lambda-u|\cdot)$ for vertical edges, whereas the left
and right boundary weights read
$$
Y(\lambda-u/2,\xi_L|r)W(\lambda-u|r) \,,
\qquad \text{and} \qquad
Y(u/2,\xi_R|r) \,.
$$
The latter coincide for the value $\xi_L=\xi_R=u/2$, where both
are proportional to $W(\lambda-u|r)$.
Hence, for this choice of boundary parameters $\xi_L,\xi_R$, the boundary
weights are equal to the bulk weights in the vertical direction, and thus 
$\eqref{eq:Y}$ corresponds to free boundary conditions for the spins
$\{\sigma_j\}$.

\section{Conclusions}

On the two examples we have studied, we have shown that the integrable
boundary weights can be obtained by imposing a simple linear condition
-- which we have called the boundary CR equation --
on the discrete contour integral
$$
\sum_P F_s(z) \delta z \,,
$$
where $P$ stands for a modified boundary face. In the case of the $\On$ model,
we have recovered the integrable boundary weights which were obtained~\cite{YungB95}
through the mapping to the 19-vertex model, whereas for the $\ZN$ model, we
have found new integrable boundary weights.
However, unlike the bulk CR equations, where the conformal spin
$s$ is generally obtained by simple consistency conditions,
it is not clear yet how to extract conformal data from
the boundary CR equations.

Our approach is quite general, and is likely to extend to any lattice model
where a solution of the bulk CR equation has been identified. Also, 
an interesting continuation of this work would be to study the boundary
CR equations in loop models with ``non-trivial'' integrable boundary
conditions, {\it i.e.} where loops touching the boundary get a different
weight~\cite{DubailJS09}.

Finally, an important possible application of our results would be to
set up proofs of convergence to variants of SLE as in~\cite{ChelkakS09,HonglerK11},
in the presence of various integrable boundary conditions.

\paragraph{\bf Acknowledegments.}
The author thanks J.L. Jacobsen and M.A. Rajabpour for stimulating
discussions, and a careful reading of the manuscript.
This work was supported by the European Research
Council (grant CONFRA 228046).


\end{document}